\documentclass[prb,aps,preprint]{revtex4-1}

\usepackage{amsmath}
\usepackage{amssymb}
\usepackage{amsbsy}
\usepackage{graphicx}
\usepackage{upgreek}

\newcommand{\unit}[2]%
{\mbox{\ensuremath{#1}}\mbox{\,\ensuremath{\mathrm{#2}}}}
\newcommand{\micro}{\upmu}
\newcommand{\YBaCuO}{\ensuremath{\mathrm{YBa_2Cu_3O_{7-\delta}}}}
\newcommand{\Jc}{\ensuremath{J_\mathrm{c}}}
\newcommand{\Jself}{\ensuremath{J_\mathrm{sf}}}
\newcommand{\JcB}{\ensuremath{\Jc(B)}}
\newcommand{\Jcd}{\ensuremath{\Jc(d)}}
\newcommand{\Jselfd}{\ensuremath{J_\mathrm{sf}(d)}}
\newcommand{\JcH}{\ensuremath{\Jc(H)}}
\newcommand{\muH}{\ensuremath{\mu_0H}}
\newcommand{\Bself}{\ensuremath{B_\mathrm{sf}}}
\newcommand{\Bf}{\ensuremath{B_\mathrm{f}}}
\newcommand{\Hf}{\ensuremath{H_\mathrm{f}}}

\begin{document}

\title{Thickness dependence of the critical current density in superconducting films:
a geometrical approach}

\author{F. Hengstberger, M. Eisterer, H. W. Weber}
\email{hengstb@ati.ac.at}
\affiliation{Atominstitut,
Vienna University of Technology,
Stadionallee 2,
1020 Vienna}

\begin{abstract}
We analyze the influence of the magnetic field
generated by the supercurrents (self-field)
on the current density distribution by numerical simulations.
The thickness of the superconducting film determines the self-field
and consequently the critical current density at zero applied field.
We find an equation,
which derives the thickness dependence of the critical current density
from its dependence on the magnetic induction.
Solutions of the equation reproduce numerical simulations to great accuracy,
thus enabling a quantification of the dependence of the self-field critical current density
with increasing film thickness.
This result is technologically relevant for the development
of coated conductors with thicker superconducting layers.
\end{abstract}

\maketitle
A decrease of the critical current density \Jc{}
with increasing film thickness $d$
was observed in thin superconducting films of \YBaCuO{}
on both single\cite{Mog90,Fol93}
and polycrystalline\cite{Fol99,Fol03,Ija06}
substrates.
Possible explanations are a degradation of the film 
or a change in the defect structure with thickness,
but also the self-field of the sample causes a significant reduction
if \Jc{} depends on the low magnetic induction $B$
generated by the transport current\cite{Ros07}.
Identifying the true cause of the decrease
is vital for the development of coated conductors,
which must be grown to higher thicknesses
to enhance their performance.

We employ numerical calculations
of the current density distribution in a thin film
to derive a practical approximation
allowing the evaluation of the self-field depression of \Jc{}
for a given \JcB{} dependence.
Therefore,
it is essential to distinguish
between the intrinsic \JcB{} of the material,
i.e.,
the (local) dependence of the critical current density
on the magnetic induction $B$ in the film,
and the average critical current density as a function of the external applied field \JcH{},
which is determined in a transport measurement of the critical current.
Note,
that  \JcB{} provides information on pinning in the material
and bridges theory and experiment.

The procedure we employ is similar to the iterative algorithm
used by Rostila et al.\cite{Ros07} 
At the beginning the external applied field determines
the starting current density distribution
$J_c(B(x,y)=\muH)$,
which is constant across the cross-section of the film.
After calculating the self-field distribution $\Bself(x,y)$ in the sample
the current densities are updated to
$J_c(B(x,y)=\muH+\Bself(x,y))$,
which results in a new self-field distribution.
This step is iterated until current density and magnetic induction
satisfy the (arbitrary) material law \JcB{} at every position in the film.
For comparison with experiment \JcH{} is computed
by calculating the average current density flowing through the conductor
at a range of external applied fields.

\begin{figure}
  \centering
  \includegraphics[width=.5\textwidth]{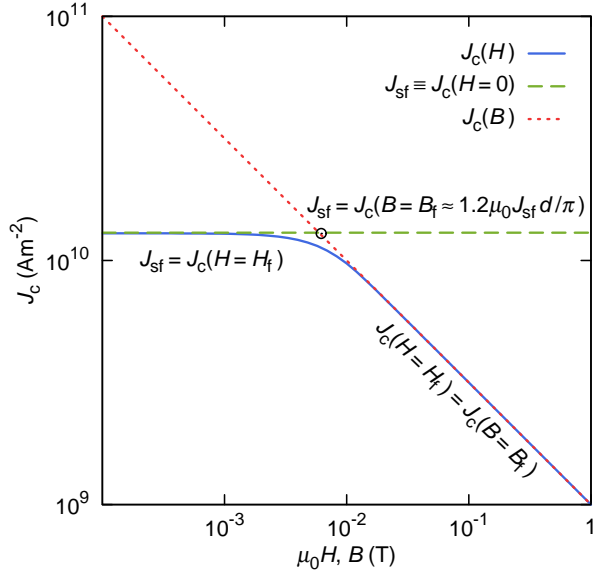}
  \caption{\label{fig:jchb}%
    Approximations made to connect \Jselfd{} and \JcB.
    The simulations of \JcH{} (solid line)
    are carried out assuming 
    $\JcB \propto B^{-0.5}$ (dotted line)
    and result in a finite \Jself{} at zero applied field.
    The simulated \JcH{} remains approximately at this value
    up to applied fields of \Bf,
    which is found by intersecting \Jself{} (broken line) and \JcB,
    and approaches \JcB{} above this field.
  }
\end{figure}

Earlier\cite{Hen09}
we used this procedure to extrapolate the power-law
$\JcB{} \propto B^{-\alpha}$
dependence,
which is commonly observed in thin films of YBCO at high fields,
to the lowest fields
and showed that despite the divergence of \JcB{}
the film carries a finite critical current at zero applied field
because of the self-field of the sample.
This result
(see Fig.~\ref{fig:jchb})
is particularly suited for our purpose,
because the clear deviation between \JcB{} and \JcH{} at low fields
emphasizes important features of the \JcH{} curve:
Starting from zero applied field,
\JcH{} remains approximately constant
at the self-field critical current density $\Jself \equiv \Jc(H=0)$
as long as the applied field is negligible compared to the self-field of the sample.
At high applied fields,
on the other hand,
the transport current alters the field distribution only marginally
and \JcH{} is identical to \JcB.
The applied field at which \JcH{} becomes field dependent and approaches \JcB{},
can be estimated by comparison to the self-field of the film.
We use
$\Hf = \Bf/\mu_0 = \gamma\Jself{}d/\pi$,
which is equivalent to the field scale of thin films\cite{Bra93},
but we replace the constant \Jc{} of the Bean model
by the average critical current density at zero applied field
and include a factor $\gamma$,
which is approximately 1.2
as determined graphically from Fig.~\ref{fig:jchb}
or by least squares fitting (see below).

Because the sheet current density $\Jself d$
controls the self-field of the sample,
the thickness dependence \Jselfd{} can be related to \JcB{}
by making the following approximations.
We neglect any field dependence of \JcH{} 
up to applied fields
$H = \Hf$,
which leads to

\begin{equation}
  \label{eqn:const}
  \Jc(H = 0)
  =
  \Jc(H = \Hf)\,,
\end{equation}

and assume that \JcH{} and \JcB{} 
are identical at applied fields above \Hf{}:

\begin{equation}
  \label{eqn:agree}
  \Jc(H = \Hf)
  =
  \Jc(B = \Bf)\,.
\end{equation}

Combining both equations

\begin{equation}
  \label{eqn:result}
  \Jc(H = 0)
  = 
  \Jc(B = \Bf)
\end{equation}

and inserting the definition of \Bf{}
results in the implicit equation

\begin{equation}
  \label{eqn:final}
  \Jself
  =
  \Jc(B = \mu_0\gamma\Jself{}d/\pi)\,,\quad\gamma\approx 1.2\,.
\end{equation}

A graphical representation of the derivation is depicted in Fig.~\ref{fig:jchb}
and $\gamma$ is determined by intersecting \Jself{} with \JcB{}.
The final result Eqn.~\ref{eqn:final} allows us to calculate
the self-field critical current density
as a function of thickness \Jselfd{} for any given \JcB{}.

\begin{figure}
  \centering
  \includegraphics[width=.5\textwidth]{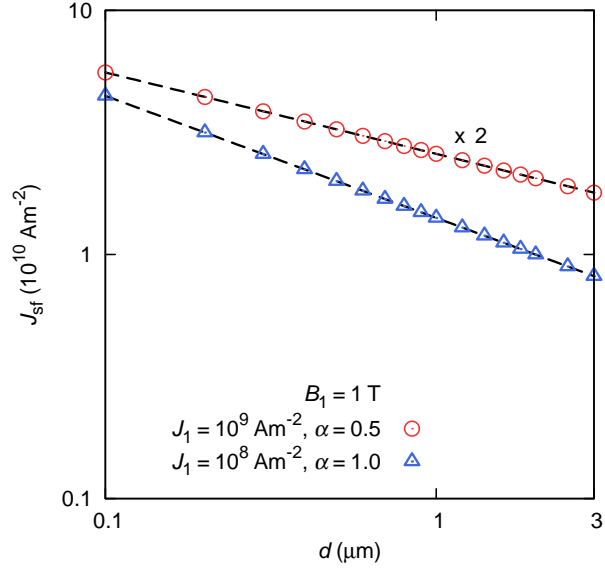}
  \caption{\label{fig:plaw}%
    Comparison of the simulated \Jselfd{} depression (points)
    to the solution of Eqn.~\ref{eqn:final} (broken lines)
    assuming a power-law \JcB.
    After fitting $\gamma \approx 1.2$
    the solution of the implicit equation
    is in excellent agreement with the numerical computation.
    (The $\alpha = 0.5$ data is shifted up by a factor of 2 for clarity.)
  }
\end{figure}

We test Eqn.~\ref{eqn:final} by comparing its solution for a certain \JcB{}
to simulations of \Jselfd{} on a \unit{100}{\micro m} wide film
having a thickness between \unit{100}{nm} and \unit{3}{\micro m}
(the typical thickness range in experiments).
Inserting,
for example,
the power-law

\begin{equation}
  \label{eqn:plaw_JcB}
  \JcB
  =
  J_1\left(\frac{B}{B_1}\right)^{-\alpha}\,,
\end{equation}

where $J_1$ is the current density at $B=B_1$,
generates a power-law also in the thickness dependence:

\begin{equation}
  \label{eqn:plaw_Jcd}
  \Jselfd
  =
  J_1^{1/(1+\alpha)} \left(\frac{\mu_0\gamma d}{B_1\pi} \right)^{-\alpha/(1+\alpha)}\,.
\end{equation}

Figure~\ref{fig:plaw} compares simulations and Eqn.~\ref{eqn:plaw_Jcd}.
We find by least squares fitting $\gamma \approx 1.2$,
in agreement with the above result.
Note further,
that this pre-factor represents only a constant vertical shift in Fig.~\ref{fig:plaw}
and that the dependence on the thickness,
which is in excellent agreement with the numerical simulations,
is thus entirely the result of Eqn.~\ref{eqn:final}.

After confirming the analytical solution for a power-law \JcB{}
we can analyze experimental data.
Fits to \Jselfd{} curves are,
for example,
available from Ijaduaola et al.\cite{Ija06},
who finds
$\Jselfd \propto d^{-0.4}$
in all but the thinnest film.
According to Eqns.~\ref{eqn:plaw_JcB} and \ref{eqn:plaw_Jcd}
the exponent $\alpha$ of \JcB{}
translates into an exponent of $\alpha/(1+\alpha)$
in the thickness dependence.
We infer
$\alpha=0.6$,
which is reasonably close to $\alpha \approx 5/8$
determined in measurements of \JcH{}
at applied fields much above the self-field,
where \JcB{} and \JcH{} are identical.
The \Jselfd{} depression observed in this work
can therefore be explained \emph{without any additional assumption}
by continuing the power-law \JcB{}
down to the low self-fields of the sample.

\begin{figure}
  \centering
  \includegraphics[width=.5\textwidth]{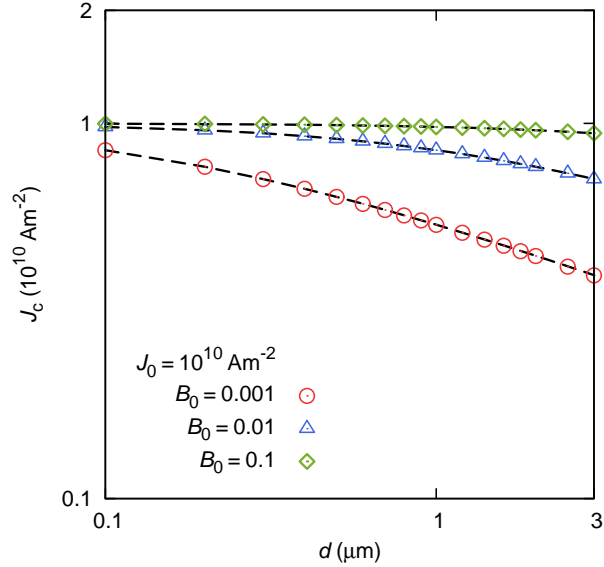}
  \caption{\label{fig:kim}
    Same as in Fig.~\ref{fig:plaw},
    but for a generalized Kim-model.
    Depending on the relative strength of $B_0$
    to the self-field of the sample,
    \Jselfd{} is almost constant if
    $B_0 \gg \Bf$ (squares)
    or decays rapidly if
    $B_0 \ll \Bf$ (circles).
    Equation~\ref{eqn:final} fully accounts for this behavior.
  }
\end{figure}

Another popular parametrization of \JcB{} is
a generalized form of the Kim model

\begin{equation}
  \label{eqn:kim_JcB}
  \JcB
  =
  \frac{J_0}{(1+B/B_0)^\alpha}\,,
\end{equation}

which reproduces a power-law if $B \gg B_0$
and is constant if $B \ll B_0$.
The maximum current density of
$\Jc(B=0)=J_0$
limits the self-field of the sample to about
$1.2 \mu_0J_0d/\pi$,
which is roughly \unit{14}{mT} at \unit{3}{\micro m}
(the thickest film of the simulations)
if we assume $J_0=\unit{10^{10}}{Am^{-2}}$.
Thus we expect that the self-field,
which increases linearly with thickness,
significantly affects \Jselfd{}
only if it is comparable to or exceeds $B_0$.
Calculations for three different values of $B_0$
are given in Fig.~\ref{fig:kim}
and show the expected behavior:
The self-field of the thinnest sample is approximately half of
$B_0=\unit{1}{mT}$
and causes a noticeable \Jc{} decrease,
whereas the self-field of the sample is insignificant
and \Jcd{} approximately independent of thickness,
if 
$B_0=\unit{100}{mT}$.
The intermediate 
$B_0=\unit{10}{mT}$
shows a transition from the first to the second case;
in the thinnest films \Jselfd{} is nearly constant 
and the depression evolves as the thickness (and the self-field) increases. 
The (numerical) solution of Eqn.~\ref{eqn:final} 
accounts for all three situations
and matches perfectly the simulations using again
$\gamma \approx 1.2$.

We demonstrated how to calculate \Jselfd{} if a theoretical \JcB{} is available,
but our result is also technologically relevant for
investigating the reason of the \Jself{} decrease
in coated conductors with thicker superconducting layers.
For such an analysis \JcH{} should be measured on the thinnest sample,
where substrate interface effects do not play a role anymore.
From this data \JcB{} is determined
down to inductions of about \Bf{} of the thinnest film.
The remaining uncertainty in \JcB{}
(we suggest to assume a constant \JcB{} at inductions below \Bf{}
if no theoretical model is available)
will introduce only a minor error in calculating \Jselfd{} for thicker films,
provided they carry higher sheet current densities
and thus elevate \Bf{} above the value of the thinnest film. 

We wish to emphasize,
that the effect of the self-field is not an alternative explanation
for a thickness dependent critical current density
and will \emph{always} be present,
if the critical current density depends on the magnetic induction.
A depression of \Jselfd{} significantly different from that calculated from \JcB{}
is thus indicative of an additional mechanism,
such as a change in the material or pinning properties with thickness,
determining the performance of the thicker conductor.
If,
however,
the calculated and the measured \Jselfd{} are similar,
the self-field controls the thickness dependence.
In this case an improvement in pinning in the entire sample
is necessary to enhance \JcB{} and to improve the conductor.

In summary we have derived a practical equation,
which bridges theory and experiment by 
relating the (local) dependence of the critical current density
on the magnetic induction \JcB{} to the thickness dependence
of the average critical current density at zero applied field \Jselfd{}.
Solutions of the equation correctly predict
the thickness dependence for two common \JcB{} models
and excellent agreement is reached by fitting a single pre-factor.
Thus the influence of \JcB{}
on the measured \Jselfd{} in transport experiments can be quantified,
which is technologically important,
because it allows investigating the depression
of the critical current density observed in coated conductors
when they are grown to higher thicknesses.


%

\end{document}